# Engineering fragile topology in photonic crystals: Topological quantum chemistry of light

María Blanco de Paz,[1] Maia G. Vergniory,[1,2] Dario Bercioux,[1,2] Aitzol García-Etxarri,[1,2,*] and Barry Bradlyn[3,†]

[1]*Donostia International Physics Center, 20018 Donostia-San Sebastian, Spain*
[2]*IKERBASQUE, Basque Foundation for Science, Maria Diaz de Haro 3, 48013 Bilbao, Spain*
[3]*Department of Physics and Institute for Condensed Matter Theory, University of Illinois at Urbana-Champaign, Urbana, Illinois 61801-3080, USA*

In recent years, there have been rapid advances in the parallel fields of electronic and photonic topological crystals. Topological photonic crystals in particular show promise for coherent transport of light and quantum information at macroscopic scales. In this work, we apply for the first time the recently developed theory of "topological quantum chemistry" to the study of band structures in photonic crystals. This method allows us to design and diagnose topological photonic band structures using only group theory and linear algebra. As an example, we focus on a family of crystals formed by elliptical rods in a triangular lattice. We show that the symmetry of Bloch states in the Brillouin zone can determine the position of the localized photonic wave packets describing groups of bands. By modifying the crystal structure and inverting bands, we show how the centers of these wave packets can be moved between different positions in the unit cell. Finally, we show that for shapes of dielectric rods, there exist isolated topological bands which do not admit a well-localized description, representing the first physical instance of "fragile" topology in a truly noninteracting system. Our work demonstrates how photonic crystals are the natural platform for the future experimental investigation of fragile topological bands.

*Introduction.* In recent years, there have been tremendous parallel advances in the fields of both topological electronic materials and engineered photonic crystals. On the one hand, topologically nontrivial band insulators have been discovered [1–6] which feature protected, gapless surface, edge, and hinge [7–13] states, as well as anomalous bulk response functions [14–16]. The interplay between topology and crystal symmetry in these phases has reinvigorated the study of band theory, and resulted in new connections between topology in momentum space and the real-space orbital structure of electronic solids [17–21]. Following Haldane and Raghu's seminal ideas [22,23], many of these concepts have also been simultaneously explored in the propagation of photons in periodic dielectric structures (photonic crystals). For instance, photonic analogs of the quantum Hall effect [24,25], quantum spin-Hall effect [26–28], quantum valley-Hall effect [29], Floquet topological insulators [30–33], mirror-Chern, and quadrupole insulating [34–42] systems have been recently discovered.

Because photons in linear dielectrics are truly noninteracting, and since they can be cheaply and easily engineered with almost any desirable lattice structure, two-dimensional photonic crystals are an ideal playground for studying topological band theory. Pioneering work in this field has primarily focused on recreating strong $\mathbb{Z}_2$ topological insulators in bosonic systems protected by crystal symmetries as a proxy for fermionic time-reversal symmetry [43,44]. Most prior work has focused on producing chiral or helical edge modes in photonic systems. However, there exist also "fragile" topological phases, which do not have protected edge states [45–50]. This fragile topology is a property of a fixed number of bands, and can be diagnosed by a bulk invariant defined in the Hilbert space spanned by those bands. A nonzero invariant reflects the inability to define exponentially localized and symmetric orbitals—Wannier functions—in real space using only the basis of states in the topological bands. With the possible exception of twisted bilayer graphene [51,52] and TMD heterostructures [53], a material realization of fragile topology without interactions has remained elusive. Systems with fragile topological bands are predicted to host tunable corner modes and defect states [13,54], which could be functionalized in a photonic crystal device. Additionally, photonic fragile topological bands can serve as the building blocks for three-dimensional photonic crystals with higher-order topology [47].

In this work, we design a photonic crystal the first known example of noninteracting fragile topology. Inspired by Ref. [43], we consider a family of photonic crystal structures with a distorted honeycomb lattice of dielectric rods. We show that by changing the shape of the dielectric rods, several different photonic band structures can be realized. To analyze these band structures, we apply the theory of topological quantum chemistry (TQC) [18] to the study of photonic band structures. In particular, we use the symmetry of the wave functions in our photonic crystals to identify the set of Wannier functions—known as a band representation– associated to each group of bands [19,55–59]. While photonic Wannier functions and band representations have been introduced previously [60–68], we here explore their topological implications for the first time. We show that there exists a class of structures where the photonic bands cannot support exponentially localized symmetric Wannier functions, and by computing the bulk invariant we verify that these bands have nontrivial fragile topology. In doing so, we also demonstrate the utility of TQC and "symmetry indicator" methods [21,52] for the efficient computation of photonic topological invariants. Finally, we comment on potential applications of TQC and fragile topology to photonic systems.

*aitzolgarcia@dipc.org
†bbradlyn@illinois.edu

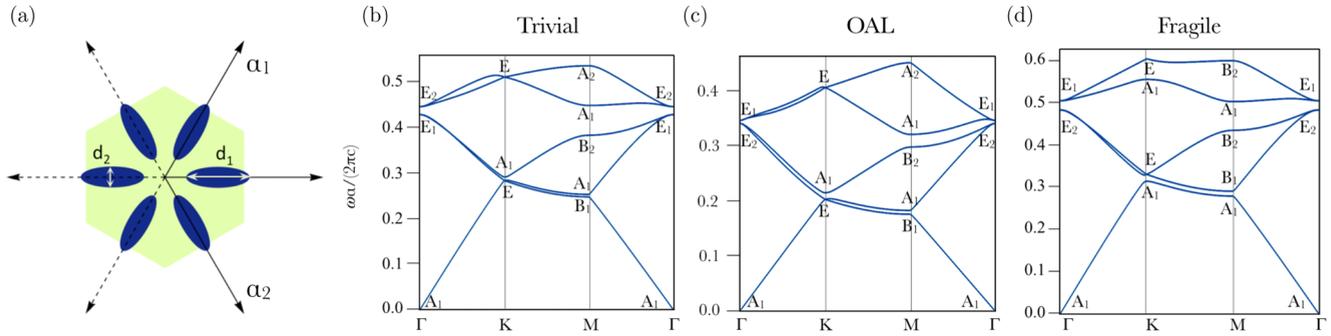

FIG. 1. (a) Schematic showing the real-space unit cell of the structures studied in this work. $a_1$ and $a_2$ are the real-space lattice vectors. The centers of the ellipses are fixed at a distance $b = \frac{a_0}{3}$ from the center of the unit cell, $a_0$ being the lattice constant. $d_1$ and $d_2$ are the lengths of the principal axes of the ellipses. Higher dielectric constant is shown in blue. When tiling this pattern we use the convention that the dielectric function in any blue region is the same, including when ellipses overlap. (b)–(d) Photonic band structures of three representative examples studied in this work, with little group representations labeled at the high-symmetry points. (b) Topologically trivial structure, with $d1 = 0.52a_0$ and $d2 = 0.31a_0$. (c) Band structure of a structure representative of the "obstructed atomic limit" (OAL) phase, with $d1 = 0.4a_0$ and $d2 = 0.61a_0$. (d) Topologically fragile structure, with $d1 = 0.4a_0$ and $d2 = 0.13a_0$.

*Model.* As a starting point for our design, we chose a two-dimensional triangular lattice of lattice constant $a_0$ with a unit cell of six circular silicon rods ($\varepsilon = 11.7$) of diameter $d$ arranged in a hexagonal pattern. This nonprimitive (enlarged) unit cell is necessary, because we next distort the rods into ellipses. These ellipses have their principal axes of length $d_1$ and $d_2$, with $d_1$ oriented in the direction of the lattice vectors [i.e., pointing towards the center of the unit cell, see Fig. 1(a)]. The symmetry group of this crystal is the space group $p6mm1'$ (No. 183) [69], generated by the lattice translations, sixfold rotational symmetry, mirror reflection about the $x$ axis, and time-reversal symmetry.

We then computed the band structure for the transverse magnetic (TM) modes in this crystal using the MIT photonic bands package (MPB) [70], given by the spectrum of the magnetic wave equation

$$\nabla \times \left(\frac{1}{\epsilon(\mathbf{r})} \times \mathbf{H}(\mathbf{r})\right) = \left(\frac{\omega}{c}\right)^2 \mathbf{H}(\mathbf{r}), \quad (1)$$

for waves with no propagation in the **z** direction. Here, $\varepsilon(\mathbf{r})$ is the position dependent permittivity, $\mathbf{H}(\mathbf{r})$ is the in-plane magnetic field vector, $\omega$ is the frequency and $c$ is the speed of light. We show representative band structures for three different cases in Fig. 1. In Fig. 2, we summarize our results as a function of the axis lengths $d_1$ and $d_2$. As we will show below, there exists a parameter regime where the second and third bands (counting up from zero energy) are isolated from the rest of the states in the spectrum and exhibit fragile topology.

*Photonic band representations.* As a first step in assessing the topological properties of bands in our photonic crystal, we will apply the theory of TQC to photonic energy bands. First, we examine the transformation properties of the Bloch eigenstates of our photonic crystal at each of the high-symmetry points $\mathbf{k}_*$ ($\Gamma$, $K$, and $M$) in the Brillouin zone. The group of symmetry operations $G_{\mathbf{k}_*}$ that leaves $\mathbf{k}_*$ invariant is known as the little group of $\mathbf{k}_*$; degenerate multiplets of states at each $\mathbf{k}_*$ transform under irreducible representations (irreps) of the group $G_{\mathbf{k}_*}$ [71]. Using the little group representations given in the Bilbao Crystallographic Server (BCS) [72–74] along with the GTPACK package [75,76], we compute the representation labels at each high-symmetry point in our photonic band structure [77]. Using these assignments, we identify three distinct phases of our model by looking at the irreps of the lowest three bands (see Table I and also Ref. [77]).

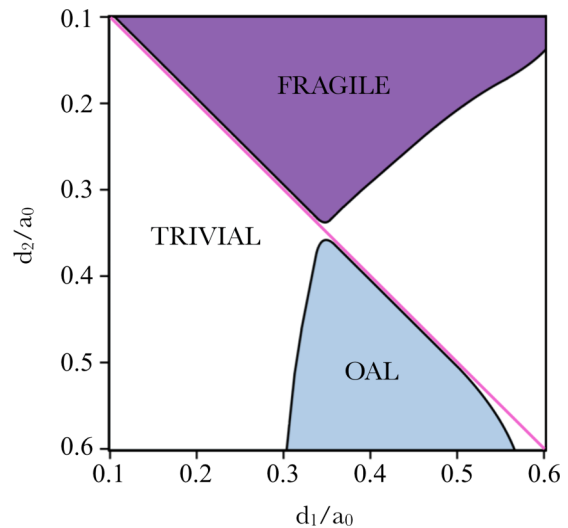

FIG. 2. Phase diagram for the photonic band topology. We show the topological properties of the second and third bands as a function of the length of principal axes of the elliptical rods. Light blue indicates that the bands form the trivial band representation, dark blue indicates the "obstructed atomic limit" (OAL), and purple indicates the fragile topological phase. Finally, magenta indicates an intervening gapless phase with fine-tuned degeneracy.

TABLE I. Little group irreps for the three gapped phases of our model. Irreps at each **k** point are ordered from lowest to highest energy. Note that while the OAL and fragile bands contain the same irreps in the lowest three bands, they differ by a band inversion at $K$ and $M$.

|         | $\Gamma$    | $K$       | $M$             |
|---------|-------------|-----------|-----------------|
| Trivial | $A_1, E_1$  | $A_1, E$  | $A_1, B_1, B_2$ |
| OAL     | $A_1, E_2$  | $E, A_1$  | $B_1, A_1, B_2$ |
| Fragile | $A_1, E_2$  | $A_1, E$  | $A_1, B_1, B_2$ |

To extract topological information from the representation labels, recall [18,59,62] that for a set of isolated bands $i \in \{1, \ldots, N\}$ the symmetry properties of the Bloch-wave eigenstates $\psi_{i\mathbf{k}}(\mathbf{r})$ at every momentum **k** in the Brillouin zone are determined by the transformation properties of the Wannier functions

$$w_{i\mathbf{R}}(\mathbf{r}) \equiv \sum_{\mathbf{k}} e^{-i\mathbf{k}\cdot\mathbf{R}} U_{ij}(\mathbf{k}) \psi_{j\mathbf{k}}(\mathbf{r}). \quad (2)$$

Here, **R** is a lattice vector, and $U_{ij}(\mathbf{k})$ is an $N \times N$ unitary matrix function of **k**, and represents a choice of "gauge" for the space spanned by the $N$ bands. For a topologically trivial set of bands, the matrix $U$ can be chosen to make the functions $w_{n\mathbf{R}}$ exponentially localized about some center $\mathbf{r}_n + \mathbf{R}$. In this case, the Wannier functions transform in a representation of the space group obtained by acting with all elements on the space group on a set of functions at one of the $\mathbf{r}_n$. These Wannier functions carry a band representation. All band representations can be obtained as a sum of elementary band representations (EBRs), which are tabulated in Refs. [78,79]. Each EBR is identified by its space group, the Wyckoff position which labels the set $\mathbf{r}_n$ of centers, and an irrep of the group $G_{\mathbf{r}_n}$ which leaves each center invariant (see SM). Inverting this observation, any set of bands that cannot be expressed as a sum of EBRs does not admit exponentially localized and symmetric Wannier functions, and is therefore topologically nontrivial. Note that these considerations apply equally well to both photonic and electronic crystals.

Using the irrep labels given in Table I, along with the catalog of EBRs in the BCS [72–74], we can identify the band representations describing each phase of our photonic crystal (see Table II of [77]). With $d_1 = 0.52a_0, d_2 = 0.31a_0$, we see that the lowest band carries irrep labels consistent with the band representation $(A_1 \uparrow G)_{1a}$, consisting of photonic Wannier functions centered at the origin with zero angular momentum ($s$-like).[1] Bands two and three are connected to each other, and are consistent with the $(E_1 \uparrow G)_{1a}$ band representation, with a pair of Wannier functions centered at the origin and transforming like a dipole ($p$-like). this is indicated as the "trivial" phase in Fig. 2, as all photonic states can be expressed in terms of modes localized near the origin. Note that there are no dielectric rods at the origin, so these Wannier functions are trapped in a symmetric arrangement of dielectrics surrounding the origin. This band structure is shown in Fig. 1(b). Next, with $d_1 = 0.4a_0, d_2 = 0.61a_0$ we see that the first three bands are all interconnected; taken together, their irrep labels are consistent with $s$-like photonic Wannier functions centered on a kagome lattice (3$c$ position), and transforming in the $(A_1 \uparrow G)_{3c}$ band representation. Note that this phase was identified in Ref. [43] as possessing a nontrivial topological invariant; here we show that this invariant indicates that the photonic Wannier functions are localized on a kagome rather than a triangular lattice. In contrast to the trivial phase, the centers of these Wannier functions lie within the dielectric rods. In analogy with similar transitions in electronic materials, this is labeled as the "obstructed atomic limit" (OAL) phase in Fig. 2. This band structure is shown in Fig. 1(c). Finally, when $d_1 = 0.4a_0, d_2 = 0.1333a_0$ we see that while the lowest band can be described by $s$-like Wannier functions at the origin of the unit cell, bands two and three cannot be expressed as the sum of EBRs. All three bands taken together, however, contain the same representations as the $(A_1 \uparrow G)_{3c}$ band representation in the OAL phase. This band structure is shown in Fig. 1(d). In the following section, we will show that bands two and three in this crystal realize fragile topology [45,46], as labeled in Fig. 2.

---

[1]The singularity near $\Gamma$ for photonic modes [90] should not obstruct the formation of Wannier functions for the electric field of TM modes in 2D, since it is polarized out of plane.

To support these conclusions, for each isolated set of bands we compute the eigenvalues of the Wilson loop

$$W = \mathcal{P} e^{i \oint \mathbf{A} \cdot d\mathbf{k}}, \quad (3)$$

where **A** is the Berry connection, $\mathcal{P}$ denotes path ordering, and the path of integration goes along a primitive reciprocal lattice vector. As shown in Ref. [77], EBRs from each of the different Wyckoff positions in this space group have qualitatively different Wilson loop spectra. Furthermore, the bands in the Wilson loop spectrum for topologically trivial bands do not cover the entire range $[0, 2\pi]$ of possible angles, i.e., they do not wind. We see in Fig. 3(a) the Wilson loop phase for the lowest band in the fragile topological phase. The phase is pinned at $\phi = 0$, consistent with a Wannier function centered at the 1$a$ position. In contrast, the Wilson loop spectrum for the second and third bands, shown in Fig. 3(b), clearly possesses nontrivial winding. The integer winding number is an indicator of nontrivial topology. That we were able to anticipated these results before performing detailed Wilson loop calculations demonstrates the utility of TQC to photonic systems.

*Fragile topology.* In contrast to a conventional topological insulator, the Wilson loop winding in the fragile topological phase is *not* a consequence of time-reversal symmetry. In fact, the crossings in the Wilson loop spectrum at $\mathbf{k}_1 = 0$ and $\mathbf{k}_1 = \pi$ are guaranteed by the twofold rotational symmetry $C_{2z}$, only due to the limited number of bands considered. Recall that the $C_{2z}$ invariant points in the Brillouin zone are $\Gamma$ and $M \equiv M' \equiv M''$. Consulting Table I, we see that the $C_{2z}$ eigenvalues of bands two and three at $\Gamma$ are $(+1, +1)$, while at all three $M$ points they are $(-1, -1)$. As was shown in Refs. [45,80], this means that the Wilson loop at $\mathbf{k}_1 = 0$, which passes through $\Gamma$ and $M$, has its phases pinned to $\pi$. Similarly, the Wilson loop at $\mathbf{k}_1 = \pi$ passes through $M'$ and $M''$, and so has its phases pinned to 0. This forces the winding of the spectrum in Fig. 3(b), and hence the nontrivial topology [77].

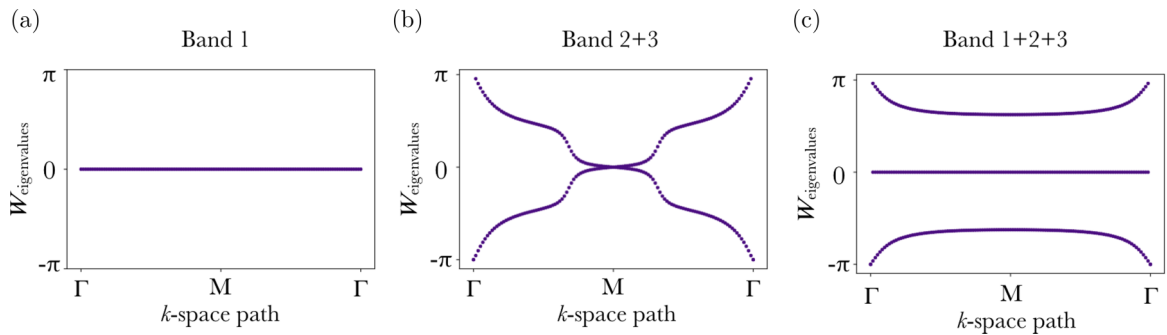

FIG. 3. Wilson loops corresponding to the lowest three in the fragile topological phase. (a) shows the Wilson loop for the isolated first band. The Wilson loop phase here is pinned to $\theta = 0$, a consequence of the $C_2$ eigenvalues of the band. (b) shows the Wilson loop eigenvalues for the interconnected second and third bands. The Wilson loop spectrum consists of two bands, which wind in opposite directions from $-\pi$ to $\pi$ as a function of momentum. As in Ref. [45], this winding is guaranteed by the $C_2$ eigenvalues of the bands, and indicates their nontrivial fragile topology. (c) shows the Wilson loop for all three bands taken together, which does not display any winding.

In fact, bands two and three realize the same fragile topological phase first discussed in a toy model in Ref. [45]. As in that case, although the irreps at the high-symmetry point in our photonic crystal do not match a sum of EBRs, they do match a difference of EBRs. In particular, we can see from the above that the irreps for bands two and three are consistent with the formal difference

$$(A_1 \uparrow G)_{3c} \ominus (A_1 \uparrow G)_{1a}, \quad (4)$$

reflecting the fact that all three bands taken together have the same irreps as in the OAL. This reflects the defining feature of fragile topology—that fragile topological bands become trivial when added to topologically trivial bands. We verify this for our model by computing the Wilson loop for all three bands, shown in Fig. 3(c). We see that the three-band Wilson loop exhibits no winding, and is consistent with Wannier functions centered at the kagome (3c) position.

*Outlook.* In this work, we have used the tools of TQC to design and characterize exotic topological photonic crystals. In particular, we have shown that the recently introduce notion of fragile topology arises naturally in photonic crystals with triangular lattice symmetry. Taking inspiration from Refs. [43,44], we have designed a family of photonic crystals with elliptical rods, which realize trivial, (fragile) topological, and "obstructed atomic limit" bands as a function of the rod shape.

To our knowledge, this represents the first controllable implementation of fragile topology. Because photonic systems are noninteracting, and individual states can be experimentally addressed, this work lays the foundation for future experimental work in probing fragile topology in photonic crystals. For instance, a variety of exotic phenomena have been predicted for fragile topological phases, such as tunable corner states and nontrivial defect modes, which could be used to develop topological lasers [81–85], as well as topological quantum optics interfaces [86–88]. Furthermore, several proposals exist for creating hinge modes in higher order topological insulators by manipulating fragile topology, which could be used for lossless information transfer [89] and signal processing [27].

By introducing TQC to the photonic domain, we highlight the importance of considering photonic Wannier functions in determining the properties of trivial *and* topological photonic systems. In particular, structures analogous to our OAL model have been confused for topologically nontrivial systems in previous works; here we show that instead the exotic properties of such systems stem from a change in position of their photonic Wanier functions. In showing how TQC can be used to design the fragile topological and obstructed atomic limit structures in this work, we provide a proof-of-principle for how TQC can be used to design and characterize photonic systems with tunable topology and tunable edge states. These tools are crucially important due to the lack of software packages for computing Wannier functions and topological invariants in photonic systems. We hope our work opens the door to a systematic study of properties of photonic Wannier functions.

*Acknowledgments.* The authors thank G. Giedke and J. J. Saenz, for initial discussions. B.B. additionally acknowledges discussions with A. Alexandradinata, B. Gadway, and I. Souza. M.G.V. and A.G.E. acknowledge the IS2016-75862-P and FIS2016-80174-P national projects of the Spanish MINECO, respectively and the OF23/2019 project of the Gipuzkoako Foru Aldundia. A.G.E. received funding from the Fellows Gipuzkoa fellowship of the Gipuzkoako Foru Aldundia through FEDER "Una Manera de hacer Europa," and µ4F (KK-2017/00089), µ4industry (KK-2019/00101) and Q-NANOFOT(IT1164-19) projects from the Basque Government. The work of M.B.d.P., D.B., and A.G.E. is supported by the Basque Government through the SOPhoQua project (Grant PI2016-41). The work of D.B. is supported by Spanish Ministerio de Ciencia, Innovation y Universidades (MICINN) under the project FIS2017-82804-P.

[1] D. J. Thouless, M. Kohmoto, M. P. Nightingale, and M. den Nijs, Phys. Rev. Lett. **49**, 405 (1982).

[2] C. L. Kane and E. J. Mele, Phys. Rev. Lett. **95**, 226801 (2005).

[3] L. Fu, C. L. Kane, and E. J. Mele, Phys. Rev. Lett. 98, 106803 (2007).

[4] B. A. Bernevig, T. L. Hughes, and S.-C. Zhang, Science 314, 1757 (2006).

[5] M. König, S. Wiedmann, C. Brüne, A. Roth, H. Buhmann, L. W. Molenkamp, X.-L. Qi, and S.-C. Zhang, Science 318, 766 (2007).

[6] Y. Xia, D. Qian, D. Hsieh, L. Wray, A. Pal, H. Lin, A. Bansil, D. Grauer, Y. S. Hor, R. J. Cava et al., Nat. Phys. 5, 398 (2009).

[7] F. Schindler, A. M. Cook, M. G. Vergniory, Z. Wang, S. S. P. Parkin, B. A. Bernevig, and T. Neupert, Sci. Adv. 4, eaat0346 (2018).

[8] E. Khalaf, H. C. Po, A. Vishwanath, and H. Watanabe, Phys. Rev. X 8, 031070 (2018).

[9] W. A. Benalcazar, B. A. Bernevig, and T. L. Hughes, Science 357, 61 (2017).

[10] W. A. Benalcazar, B. A. Bernevig, and T. L. Hughes, Phys. Rev. B 96, 245115 (2017).

[11] W. A. Benalcazar, T. Li, and T. L. Hughes, Phys. Rev. B 99, 245151 (2019).

[12] Z. Song, Z. Fang, and C. Fang, Phys. Rev. Lett. 119, 246402 (2017).

[13] B. J. Wieder, Z. Wang, J. Cano, X. Dai, L. M. Schoop, B. Bradlyn, and B. A. Bernevig, arXiv:1908.00016.

[14] X.-L. Qi, T. L. Hughes, and S.-C. Zhang, Phys. Rev. B 78, 195424 (2008).

[15] A. M. Essin, J. E. Moore, and D. Vanderbilt, Phys. Rev. Lett. 102, 146805 (2009).

[16] J. C. Y. Teo and C. L. Kane, Phys. Rev. B 82, 115120 (2010).

[17] A. A. Soluyanov and D. Vanderbilt, Phys. Rev. B 83, 035108 (2011).

[18] B. Bradlyn, L. Elcoro, J. Cano, M. G. Vergniory, Z. Wang, C. Felser, M. I. Aroyo, and B. A. Bernevig, Nature (London) 547, 298 (2017).

[19] J. Cano, B. Bradlyn, Z. Wang, L. Elcoro, M. G. Vergniory, C. Felser, M. I. Aroyo, and B. A. Bernevig, Phys. Rev. B 97, 035139 (2018).

[20] J. Höller and A. Alexandradinata, Phys. Rev. B 98, 024310 (2018).

[21] H. C. Po, A. Vishwanath, and H. Watanabe, Nat. Commun. 8, 50 (2017).

[22] F. D. M. Haldane and S. Raghu, Phys. Rev. Lett. 100, 013904 (2008).

[23] S. Raghu and F. D. M. Haldane, Phys. Rev. A 78, 033834 (2008).

[24] Z. Wang, Y. D. Chong, J. D. Joannopoulos, and M. Soljačić, Phys. Rev. Lett. 100, 013905 (2008).

[25] Z. Wang, Y. Chong, J. D. Joannopoulos, and M. Soljačić, Nature (London) 461, 772 (2009).

[26] A. B. Khanikaev, S. H. Mousavi, W.-K. Tse, M. Kargarian, A. H. MacDonald, and G. Shvets, Nat. Mater. 12, 233 (2013).

[27] M. Hafezi, E. A. Demler, M. D. Lukin, and J. M. Taylor, Nat. Phys. 7, 907 (2011).

[28] R. O. Umucalılar and I. Carusotto, Phys. Rev. A 84, 043804 (2011).

[29] T. Ma and G. Shvets, New J. Phys. 18, 025012 (2016).

[30] K. Fang, Z. Yu, and S. Fan, Nat. Photonics 6, 782 (2012).

[31] L. J. Maczewsky, J. M. Zeuner, S. Nolte, and A. Szameit, Nat. Commun. 8, 13756 (2017).

[32] S. Mukherjee, A. Spracklen, M. Valiente, E. Andersson, P. Öhberg, N. Goldman, and R. R. Thomson, Nat. Commun. 8, 13918 (2017).

[33] M. C. Rechtsman, J. M. Zeuner, Y. Plotnik, Y. Lumer, D. Podolsky, F. Dreisow, S. Nolte, M. Segev, and A. Szameit, Nature (London) 496, 196 (2013).

[34] L. Zhang, Y. Yang, P. Qin, Q. Chen, F. Gao, E. Li, J.-H. Jiang, B. Zhang, and H. Chen, arXiv:1901.07154 [physics.app-ph].

[35] A. E. Hassan, F. K. Kunst, A. Moritz, G. Andler, E. J. Bergholtz, and M. Bourennane, Nat. Photon. 13, 697 (2019).

[36] Y. Ota, F. Liu, R. Katsumi, K. Watanabe, K. Wakabayashi, Y. Arakawa, and S. Iwamoto, Optica 6, 786 (2019).

[37] B. Y. Xie, H. F. Wang, X. Y. Zhu, M. H. Lu, and Y. F. Chen, Phys. Rev. B 98, 205147 (2018).

[38] L. Lu, C. Fang, L. Fu, S. G. Johnson, J. D. Joannopoulos, and M. Soljačić, Nat. Phys. 12, 337 (2016).

[39] L. J. Maczewsky, B. Höckendorf, M. Kremer, T. Biesenthal, M. Heinrich, A. Alvermann, H. Fehske, and A. Szameit, arXiv:1812.07930.

[40] T. Ozawa, H. M. Price, A. Amo, N. Goldman, M. Hafezi, L. Lu, M. Rechtsman, D. Schuster, J. Simon, O. Zilberberg et al., Rev. Mod. Phys. 91, 015006 (2019).

[41] A. Slobozhanyuk, S. H. Mousavi, X. Ni, D. Smirnova, Y. S. Kivshar, and A. B. Khanikaev, Nat. Photonics 11, 130 (2017).

[42] L. Lu, J. D. Joannopoulos, and M. Soljačić, Nat. Photonics 8, 821 (2014).

[43] L.-H. Wu and X. Hu, Phys. Rev. Lett. 114, 223901 (2015).

[44] G. Siroki, P. A. Huidobro, and V. Giannini, Phys. Rev. B 96, 041408(R) (2017).

[45] J. Cano, B. Bradlyn, Z. Wang, L. Elcoro, M. G. Vergniory, C. Felser, M. I. Aroyo, and B. A. Bernevig, Phys. Rev. Lett. 120, 266401 (2018).

[46] H. C. Po, H. Watanabe, and A. Vishwanath, Phys. Rev. Lett. 121, 126402 (2018).

[47] B. J. Wieder and B. A. Bernevig, arXiv:1810.02373.

[48] J. Ahn and B.-J. Yang, Phys. Rev. B 99, 235125 (2019)

[49] A. Bouhon, A. M. Black-Schaffer, and R.-J. Slager, arXiv:1804.09719.

[50] J. L. Mañes, arXiv:1904.06997.

[51] L. Zou, H. C. Po, A. Vishwanath, and T. Senthil, Phys. Rev. B 98, 085435 (2018).

[52] Z. Song, Z. Wang, W. Shi, G. Li, C. Fang, and B. A. Bernevig, Phys. Rev. Lett. 123, 036401 (2019).

[53] M. P. Zaletel and J. Y. Khoo, arXiv:1901.01294.

[54] S. Liu, A. Vishwanath, and E. Khalaf, Phys. Rev. X 9, 031003 (2019).

[55] J. Zak, Phys. Rev. Lett. 45, 1025 (1980).

[56] J. Zak, Phys. Rev. B 23, 2824 (1981).

[57] J. Zak, Phys. Rev. B 26, 3010 (1982).

[58] H. Bacry, L. Michel, and J. Zak, Symmetry and classification of energy bands in crystals, in *Group Theoretical Methods in Physics: Proceedings of the XVI International Colloquium Held at Varna, Bulgaria, June 15–20 1987* (Springer, Berlin, Heidelberg, 1988), p. 289.

[59] L. Michel and J. Zak, Phys. Rep. 341, 377 (2001).

[60] K. M. Leung, JOSA B 10, 303 (1993).

[61] K. Busch, S. F. Mingaleev, A. Garcia-Martin, M. Schillinger, and D. Hermann, J. Phys.: Condens. Matter 15, R1233 (2003).

[62] K. Busch, C. Blum, A. M. Graham, D. Hermann, M. Köhl, P. Mack, and C. Wolff, J. Mod. Opt. 58, 365 (2011).

[63] C. Wolff and K. Busch, Physica B **407**, 4051 (2012).

[64] C. Wolff, P. Mack, and K. Busch, Phys. Rev. B **88**, 075201 (2013).

[65] D. M. Whittaker and M. P. Croucher, Phys. Rev. B **67**, 085204 (2003).

[66] J. P. Albert, C. Jouanin, D. Cassagne, and D. Bertho, Phys. Rev. B **61**, 4381 (2000).

[67] J. Albert, C. Jouanin, D. Cassagne, and D. Monge, Opt. Quantum Electron. **34**, 251 (2002).

[68] M. L. V. D'yerville, D. Monge, D. Cassagne, and J. Albert, Opt. Quantum Electron. **34**, 445 (2002).

[69] M. I. Aroyo (ed.), *International Tables for Crystallography*, Vol. A: Space-Group Symmetry (John Wiley and Sons, New York, 2016).

[70] S. G. Johnson and J. D. Joannopoulos, Opt. Express **8**, 173 (2001).

[71] C. J. Bradley and A. P. Cracknell, *The Mathematical Theory of Symmetry in Solids* (Clarendon Press, Oxford, 1972).

[72] M. I. Aroyo, J. M. Perez-Mato, D. Orobengoa, E. Tasci, G. de la Flor, and A. Kirov, Bulg. Chem. Commun. **43**, 183 (2011).

[73] M. I. Aroyo, J. M. Perez-Mato, C. Capillas, E. Kroumova, S. Ivantchev, G. Madariaga, A. Kirov, and H. Wondratschek, Z. Kristallogr. **221**, 15 (2006).

[74] M. I. Aroyo, A. Kirov, C. Capillas, J. M. Perez-Mato, and H. Wondratschek, Acta Crystallogr. **A62**, 115 (2006).

[75] R. M. Geilhufe and W. Hergert, Frontiers Phys. **6**, 86 (2018).

[76] W. Hergert and R. M. Geilhufe, *Group Theory in Solid State Physics and Photonics: Problem Solving with Mathematica* (Wiley-VCH, Weinheim, 2018).

[77] See Supplemental Material at http://link.aps.org/supplemental/10.1103/PhysRevResearch.1.032005 for properties and computational details of Wilson loops used in the main text.

[78] Bilbao Crystallogr. Server, "Bandrep: Band representations of the double space groups" (2017), http://www.cryst.ehu.es/cgi-bin/cryst/programs/bandrep.pl.

[79] L. Elcoro, B. Bradlyn, Z. Wang, M. G. Vergniory, J. Cano, C. Felser, B. A. Bernevig, D. Orobengoa, G. de la Flor, and M. I. Aroyo, J. Appl. Crystallogr. **50**, 1457 (2017).

[80] A. Alexandradinata, X. Dai, and B. A. Bernevig, Phys. Rev. B **89**, 155114 (2014).

[81] P. St-Jean, V. Goblot, E. Galopin, A. Lemaître, T. Ozawa, L. Le Gratiet, I. Sagnes, J. Bloch, and A. Amo, Nat. Photonics **11**, 651 (2017).

[82] B. Bahari, A. Ndao, F. Vallini, A. El Amili, Y. Fainman, and B. Kanté, Science **358**, 636 (2017).

[83] A. Kodigala, T. Lepetit, Q. Gu, B. Bahari, Y. Fainman, and B. Kanté, Nature (London) **541**, 196 (2017).

[84] G. Harari, M. A. Bandres, Y. Lumer, M. C. Rechtsman, Y. D. Chong, M. Khajavikhan, D. N. Christodoulides, and M. Segev, Science **359**, eaar4003 (2018).

[85] M. A. Bandres, S. Wittek, G. Harari, M. Parto, J. Ren, M. Segev, D. N. Christodoulides, and M. Khajavikhan, Science **359**, eaar4005 (2018).

[86] S. Barik, A. Karasahin, C. Flower, T. Cai, H. Miyake, W. DeGottardi, M. Hafezi, and E. Waks, Science **359**, 666 (2018).

[87] P. Lodahl, S. Mahmoodian, S. Stobbe, A. Rauschenbeutel, P. Schneeweiss, J. Volz, H. Pichler, and P. Zoller, Nature (London) **541**, 473 (2017).

[88] P. M. Vora, A. S. Bracker, S. G. Carter, M. Kim, C. S. Kim, and D. Gammon, Phys. Rev. B **99**, 165420 (2019).

[89] M. I. Shalaev, W. Walasik, A. Tsukernik, Y. Xu, and N. M. Litchinitser, Nat. Nanotech. **14**, 31 (2019).

[90] H. Watanabe and L. Lu, Phys. Rev. Lett. **121**, 263903 (2018).